# Photoemission study of the electronic structure and charge density waves of $Na_2Ti_2Sb_2O$

S. Y. Tan[1,2], J. Jiang[2,3], Z. R. Ye[2], X. H. Niu[2,3], Y. Song[4], C. L. Zhang[4,5], P. C. Dai[4], B. P. Xie[2,3], X. C. Lai[1] and D. L. Feng[2,3]*

[1] Science and Technology on Surface Physics and Chemistry Laboratory, Mianyang 621907, China

[2] Physics Department, Applied Surface Physics State Key Laboratory, and Advanced Materials Laboratory, Fudan University, Shanghai 200433, China

[3] Collaborative Innovation Center of Advanced Microstructures, Nanjing University, Nanjing 210093, China

[4] Department of Physics and Astronomy, Rice University, Houston, Texas 77005, USA

[5] Department of Physics and Astronomy, The University of Tennessee, Knoxville, Tennessee 37996-1200, USA

*Correspondence and requests for materials should be addressed to D.L.Feng(dlfeng@fudan.edu.cn)

**The electronic structure of $Na_2Ti_2Sb_2O$, a parent compound of the newly discovered titanium-based oxypnictide superconductors, is studied by photon energy and polarization dependent angle-resolved photoemission spectroscopy (ARPES). The obtained band structure and Fermi surface agree well with the band structure calculation of $Na_2Ti_2Sb_2O$ in the non-magnetic state, which indicating that there is no magnetic order in $Na_2Ti_2Sb_2O$ and the electronic correlation is weak. Polarization dependent ARPES results suggest the multi-band and multi-orbital nature of $Na_2Ti_2Sb_2O$. Photon energy dependent ARPES results suggest that the electronic structure of $Na_2Ti_2Sb_2O$ is rather two-dimensional. Moreover, we find a density wave energy gap forms below the transition temperature and reaches 65 meV at 7 K, indicating that $Na_2Ti_2Sb_2O$ is likely a weakly correlated CDW material in the strong electron-phonon interaction regime.**

Layered compounds of transition-metal elements always show interesting and novel electric and magnetic properties and have been studied extensively. The discovery of basic superconducting layers, such as the $CuO_2$ plane[1] in cuprates and $Fe_2An_2$ (An = P, As, S, Se, Te) layers[2] in iron based superconductors, have opened new fields in physics and chemistry of layered superconductors. Recently another class of layered compounds built from alternatively stacking of special conducting octahedral layers $Ti_2Pn_2O$ (Pn=Sb, As) and certain charge reservoir layers [e.g., $Na_2$, Ba, $(SrF)_2$, $(SmO)_2$] have attracted much attention[3-17]. Most notably, these compounds exhibit competing phases just like in cuprates and iron based superconductors. Both experiments and band calculations show that the ground states of $Na_2Ti_2Sb_2O$ (Ref. 6, 9, 18-19) and $BaTi_2Sb_2O$ (Ref. 12, 13, 20, 21) are possible spin-density wave (SDW) or charge-density wave (CDW) phases, and the $Na^+$ substitution of $Ba^{2+}$ in $Na_xBa_{1-x}Ti_2Sb_2O$ suppresses the CDW/SDW, and leads to superconductivity, whose critical temperature ($T$c) can be as high as 5.5 K for x=0.15(Ref. 13). These layered compounds provide a new platform to study unconventional superconductivity.

    $Na_2Ti_2Sb_2O$ is a sister compound to $BaTi_2Sb_2O$, which shows a phase transition at $T$s~115 K as characterized by a sharp jump in resistivity and a drop in spin susceptibility[3]. The microscopic mechanism for this phase transition has not been determined, but it has been suggested to arise from the SDW or CDW instability driven by the strongly nested electron and hole Fermi surfaces

(Ref.18-23). However, the nature of the phase transition and its correlation with the superconductivity are still unknown. A recent DFT calculation[23] predicted possible SDW instabilities in $Na_2Ti_2Pn_2O$ (Pn=As, Sb), and more specifically that the ground states of $Na_2Ti_2Sb_2O$ and $Na_2Ti_2As_2O$ are bi-collinear antiferromagnetic semimetal and novel blocked checkerboard antiferromagnetic semiconductor, respectively. An optical study[24] reveals a significant spectral change across the phase transition and the formation of a density-wave-like energy gap. However, one cannot distinguish whether the ordered state is CDW or SDW since both states have the same coherent factor. To date, the experimental electronic structure of $Na_2Ti_2Sb_2O$ has not been reported, which is critical for understanding the nature of the density waves in these compounds.

In this article, we investigate the electronic structure of $Na_2Ti_2Sb_2O$ with angle-resolved photoemission spectroscopy (ARPES). Our polarization and photon energy dependent studies reveal the multi-orbital and weak three-dimensional nature of this material. The obtained band structure and Fermi surface agree well with the band structure calculation of $Na_2Ti_2Sb_2O$ in the non-magnetic state, which indicating that there is no magnetic order in $Na_2Ti_2Sb_2O$ and the electronic correlation is weak. Temperature dependent ARPES results reveal that a density wave energy gap forms below the transition temperature and reaches 65 meV at 7 K, indicating that $Na_2Ti_2Sb_2O$ is likely a weakly correlated CDW material in the strong electron-phonon interaction regime.

## Results

**Band Structure.** The electronic structure of $Na_2Ti_2Sb_2O$ at 15 K is presented in Fig.1. Photoemission intensity maps are integrated over a [$E_F$-10 meV, $E_F$+10meV] window around the Fermi energy ($E_F$) as shown in Figs.1(a) and 1(b). The azimuth angle of the sample in Fig.1(b) was rotated by 45° compared with in Fig.1(a), there is subtle spectrum weight difference in the two obtained Fermi surface maps due to the matrix element effect. The observed Fermi surface consists of four square-shaped hole pockets (α) centered at X and four similar electron pockets (γ) centered at M. The electronic structure around Γ is more complicated, mainly consists of a diamond-shaped electron pocket (β) and four strong points (or lines, β′) along the Γ-M direction. The extracted Fermi surface from photoemission intensity map and the theoretic predicted Fermi surface are shown in Figs.1(c) and 1(d), which agree well with each other. The calculated Fermi surface of $Na_2Ti_2Sb_2O$ in the non-magnetic state was taken from Ref. 23. The Fermi pockets centered at X and M show multiple parallel sections, providing possible Fermi surface nesting condition for density wave instabilities, as suggested in previous first principle calculations [23].

The valence band structures of $Na_2Ti_2Sb_2O$ along Γ-M and Γ-X are present in Figs.1(e1) and 1(f1). The valence band structures agree qualitatively well with the calculations[23] in non-magnetic state (Figs.1(e2) and 1(f2)). Taking two distinct bands δ and η as examples, the renormalization factors are about 1~2 for both bands, suggesting the weak correlation character of $Na_2Ti_2Sb_2O$. Figs.1(g) and (h) show the low energy electronic structure along the Γ-M and Γ-X directions together with their second derivative spectrum. The band structure as indicated by the dashed curves in Figs.1(g2) and (h2) are resolved by tracking the local minimum locus in the second derivative of the ARPES intensity plot with respect to energy. A weak but dispersive electron band can be resolved around M point, its band bottom locate at the top of a hole-like band δ. Two nearly coincident electron-like bands(β and β′) can be resolved around Γ point at certain photon energy along the Γ-M direction, while there is only one electron-like band β across $E_F$ near Γ along the Γ-X direction. A hole-like band α crosses $E_F$ and forms the square-shaped pockets around X. The overall measured electronic structure of $Na_2Ti_2Sb_2O$ agrees well with the calculations, and the near-unity renormalization factor suggests that the ground state of $Na_2Ti_2Sb_2O$ is nonmagnetic and the correlation is weak.

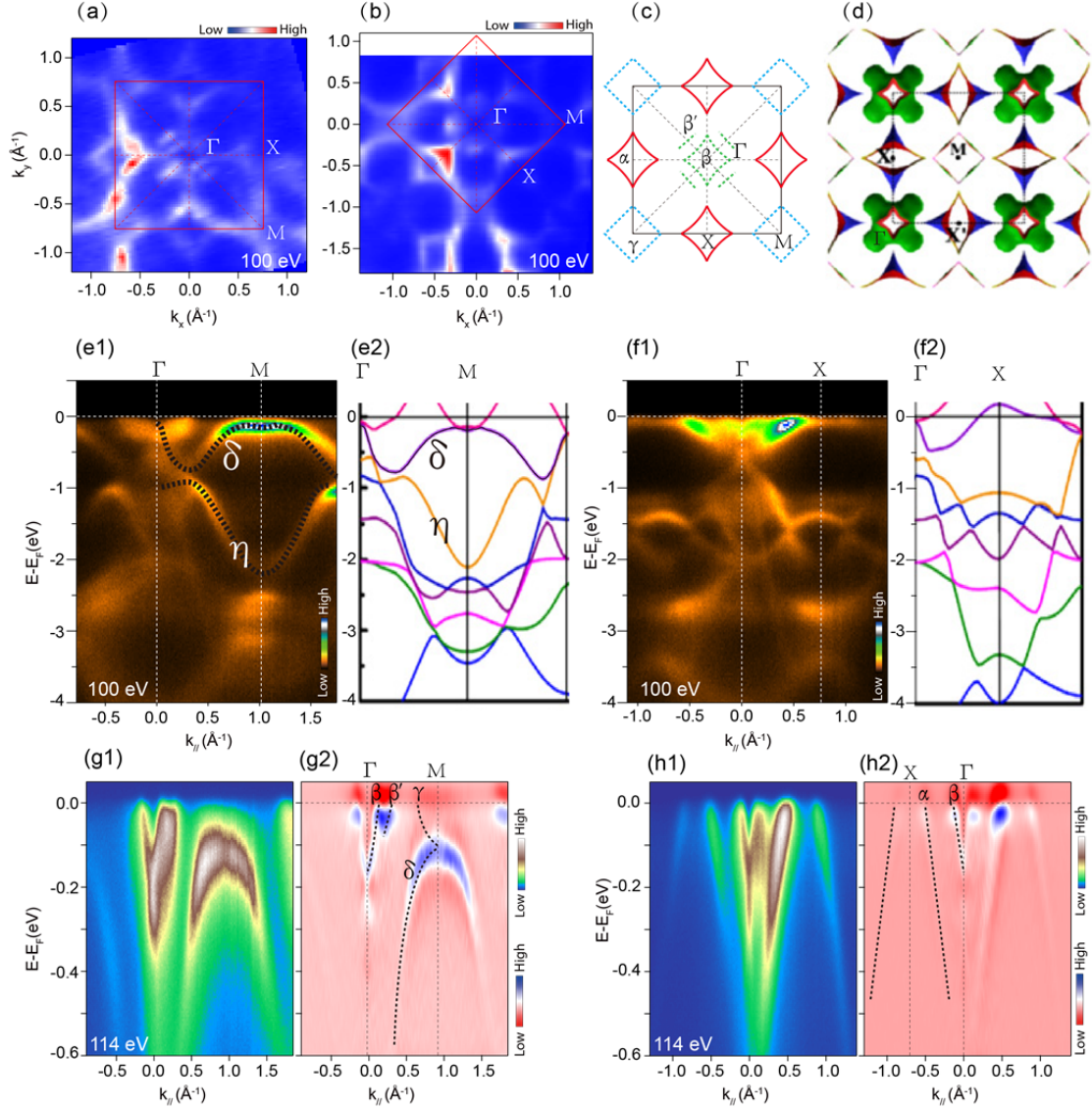

**Fig.1** The electronic structure of $Na_2Ti_2Sb_2O$ at 15 K. (a) and (b), Photoemission intensity map at $E_F$ integrated over [$E_F$ -10 meV, $E_F$ +10 meV], data were measured with 100 eV photons energy. (c), The Fermi surface topology extract from (a) and (b). (d), The theoretic predicted Fermi surface of $Na_2Ti_2Sb_2O$. (e1-e2), The experiment and theoretic[23] valence band structure along Γ-M direction. (f1-f2), The experiment and theoretic[23] valence band structure along Γ-X direction. Data were measured with 100 eV photons energy in (e) and (f). (g1-g2), The photoemission intensity and its second derivative of the intensity plot with respect to energy along Γ-M direction. (h1-h2), The photoemission intensity and its second derivative of the intensity plot with respect to energy along Γ-X direction, data were measured with 114 eV photons energy.

**Polarization Dependence.** The electronic structure of $Na_2Ti_2Sb_2O$ near $E_F$ is mainly contributed by Ti $3d$ orbitals, which is similar to the case of iron based superconductors. We conducted the polarization dependent photoemission spectroscopy measurement to resolve the possible multi-orbital nature of $Na_2Ti_2Sb_2O$. The experimental setup for polarization-dependent ARPES is shown in Fig. 2(a). The incident beam and the sample surface normal define a mirror plane. For the $s$ (or $p$) experimental geometries, the electric field of the incident photons is out of (or in) the mirror plane. The matrix element for the photoemission process could be described as:

$$M_{f,i}^k \propto \left|\langle \Psi_f^k | \hat{\varepsilon} \cdot r | \Psi_i^k \rangle\right|$$

Since the final state $\Psi_f^k$ of photoelectrons could be approximated by a plane wave with its wave vector in the mirror plane, is always even with respect to the mirror plane in our experimental geometry. In the $s$ (or $p$) geometry, $\hat{\varepsilon} \cdot r$ is odd (or even) with respect to the mirror plane. Thus considering the spatial symmetry of the Ti $3d$ orbitals, when the analyzer slit is along the high-symmetry directions, the photoemission intensity of specific even (or odd) component of a band is only detectable with the $p$ (or $s$) polarized light. For example, with respect to the mirror plane (the $xz$ plane), the even orbitals ($d_{xz}$, $d_{z^2}$, and $d_{x^2-y^2}$) and the odd orbitals ($d_{xy}$ and $d_{yz}$) could be only observed in the $p$ and $s$ geometries, respectively.

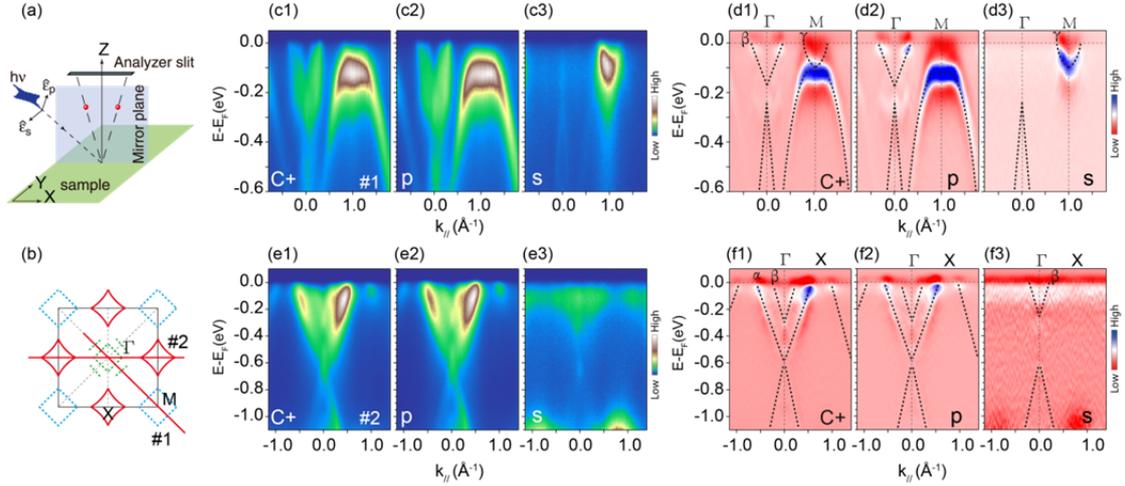

**Fig.2** Polarization dependent photoemission data of $Na_2Ti_2Sb_2O$ at 15 K. (a), Experimental setup for polarization-dependent ARPES. (b), The Brillouin zone of $Na_2Ti_2Sb_2O$ and locations of the momentum cuts. (c1-c3) and (d1-d3), The photoemission intensity and its second derivative of the intensity plot with respect to energy along Γ-M taken with $C+$, $p$ and $s$ polarized light, respectively. (e1-e3) and (f1-f3), The photoemission intensity and its second derivative of the intensity plot with respect to energy along Γ-X taken with $C+$, $p$ and $s$ polarized light, respectively, data were measured with 100 eV photons energy.

The photoemission intensity plots of $Na_2Ti_2Sb_2O$ along the Γ-M and Γ-X high symmetry directions are shown in Fig.2. The incident $C+$ light is a mixture of both the $p$ and $s$ polarizations, so all the bands with specific orbital can be seen with the $C+$ incident light. The β band at Γ is absent in the $s$ polarization along the Γ-M direction, visible in both polarizations along the Γ-X direction, which may be attributed to the Ti $d_{xz}$ orbital. The electron band γ only shows up on the $s$ polarization at the M point, exhibiting its odd nature with respect to the mirror plane, which may be attributed to the $d_{yz}$ and/or $d_{xy}$ orbital. The hole-like band at X point is not as pure, it is visible in the $p$ polarization along the Γ-X direction, hardly seen in the $s$ polarization, which may be a mixture of different Ti 3*d* orbitals. In general, Na$_2$Ti$_2$Sb$_2$O exhibits obvious polarization dependence, which resembles the multi-band and multi-orbital nature of band structure of iron pnictide superconductors[25].

**Kz Dependence.** The calculated electronic structure of Na$_2$Ti$_2$Sb$_2$O shows typical two dimensional character by the nearly *kz*-independent Fermi surface sheets around the X and M points, while the electronic structure exhibit significant $k_z$ dispersion at Γ point[22,23]. To study the three-dimensional character of the electronic structure in Na$_2$Ti$_2$Sb$_2$O, we have conducted the photon energy dependent experiment with circularly polarized photons. The measured band structures along the two high-symmetry directions (Γ-M and Γ-X) with different photon energies are present in Fig.3. Figs. 3(a) and (d) show the band dispersion and Fermi crossing along the Γ-X direction, where an electron-like band marked as β and a hole-like band marked as α cross the Fermi energy. The Fermi crossings of bands are determined by tracking the peak positions in the MDCs taken at various photon energies. The Fermi crossings of α and β bands both show weak $k_z$ dispersion with a typical cycle of each 14 eV photon energy. The Fermi momentum of β reaches its minimal at 90 eV photon energy, then increases with increasing photon energy, and reaches its maximal at 104 eV. On the contrary, the Fermi momentum of α band reaches its maximal and minimal at 90 eV and 104 eV, respectively.

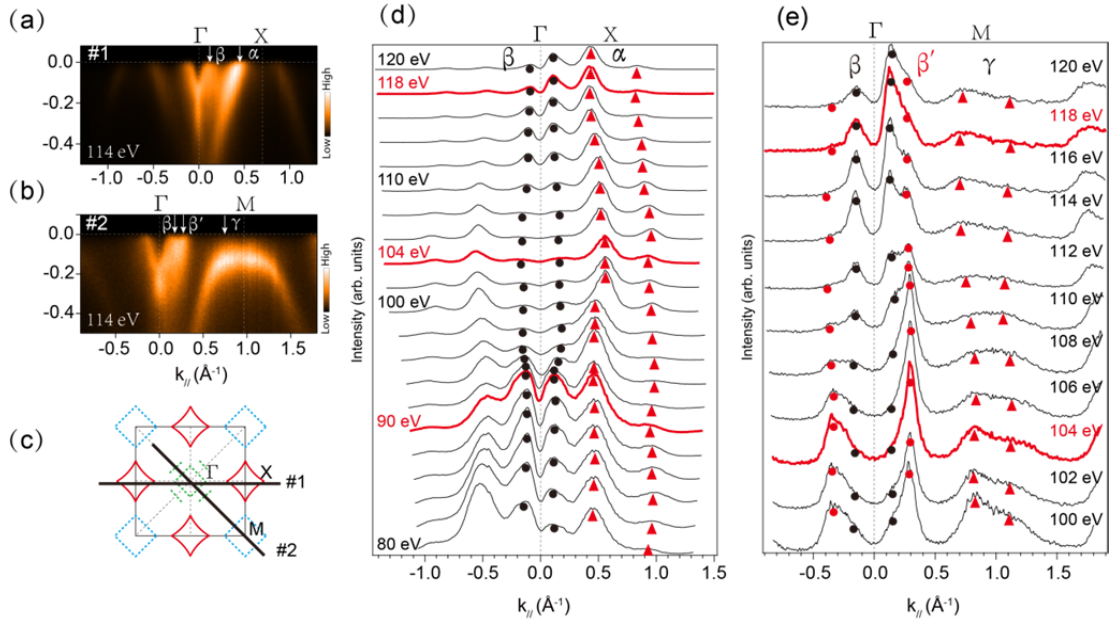

**Fig.3** Photon energy dependence of the band structure of Na$_2$Ti$_2$Sb$_2$O at 15 K. (a), Photoemission intensity along Γ-X taken with 114 eV photons[cut #1 in (c)]. (b), Photoemission intensity along Γ-M with 114 eV photon [cut #2 in (c)]. (c), The Brillouin zone of Na$_2$Ti$_2$Sb$_2$O and the experimental momentum cuts. (d), Photon energy dependence of the MDCs along Γ-X. (e), Photon energy dependence of the MDCs along Γ-M.

Consistent with the measured Fermi surface, there is only one electron band near Γ along the Γ-X direction (labeled as β), while we can clearly observe two electron bands along the Γ-M direction (labeled as β and β′). The Fermi crossings of β and β′ show negligible photon energy dependence along Γ-M, while the relative intensity of β and β′ change with photon energy. For instance at 104 eV, the β′ intensity is high, while the β intensity is low. With increasing photon energy, the intensity of β′ decreases while that of β increases, reaching their minimum and maximum at 118 eV, respectively. The relative intensity instead of Fermi crossing shows distinct photon energy dependence for β and β′. For the γ band near the M point, its Fermi momentum shows weak kz dispersion, with the minimal and maximum at 104 eV and 118 eV, respectively.

The theoretic predicted Fermi surface of $Na_2Ti_2Sb_2O$ shows cylinder Fermi sheets near M and X and strong $k_z$ dependent Fermi sheet near Γ[22,23], our photoemission data confirmed the two dimensional character of the electronic structure at X and M. The weak photon energy dependence of the electronic structure at Γ is not consistent with the theoretic calculation, and this discrepancy may be due to the poor $k_z$ resolution of our ARPES experiment in the vacuum ultra-violet photon energy range. It is known that the poor $k_z$ resolution would largely smear out the dispersive information along $k_z$ for a fast-dispersive band, as likely observed here.

**Formation of the Density Wave Energy Gap.** In the conventional picture of density wave transition, the formation of electron-hole pairs with a nesting wave vector connecting different regions of FSs would lead to the opening of an energy gap. In charge-density wave systems such as $2H-TaS_2$, strong electron–photon interactions could cause incoherent polaronic spectral lineshape, and large Fermi patches instead of a clear-cut Fermi surface[26]. Anomalous temperature dependent spectral weight redistribution and broad lineshape with incoherent character was reported in $BaTi_2As_2O$ (Ref. 27), an iso-structural compound of $Na_2Ti_2Sb_2O$. It was found that partial energy gap opens at the Fermi patches, instead of Fermi surface nesting, is responsible for the CDW in $BaTi_2As_2O$.

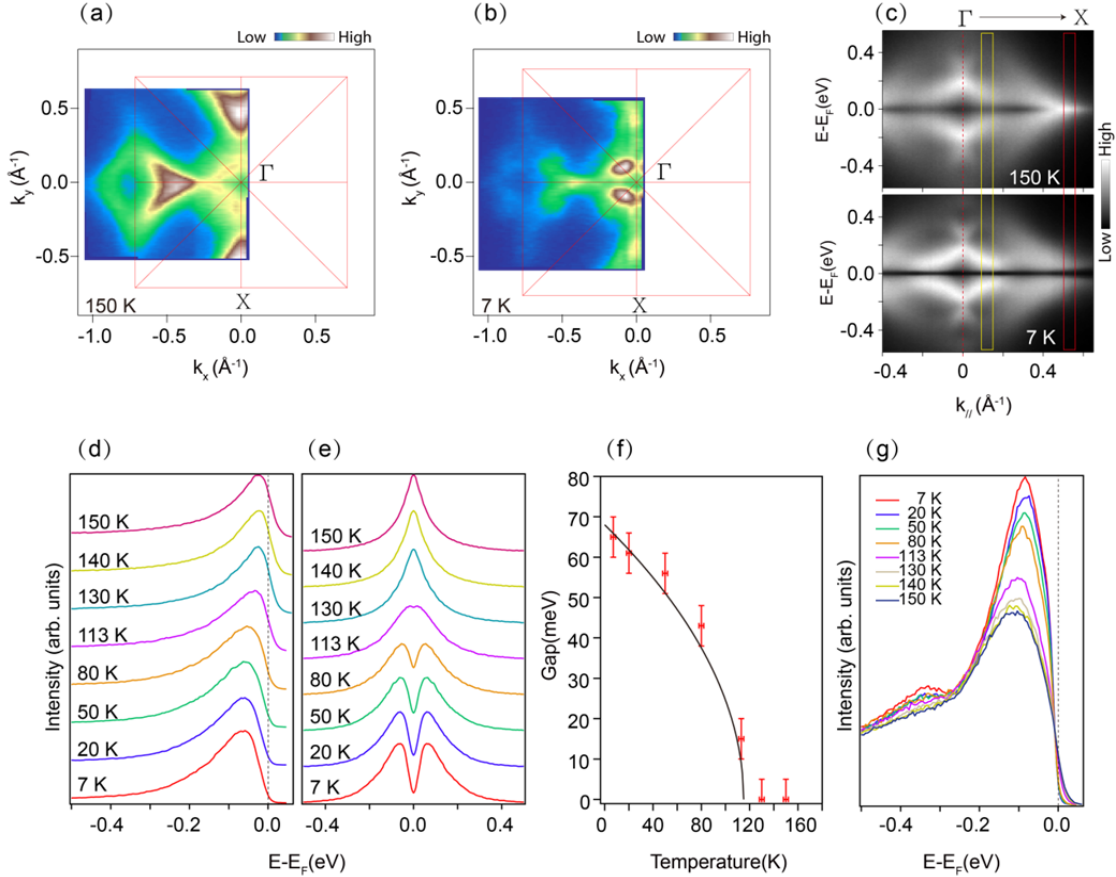

**Fig.4** The temperature dependence of Na$_2$Ti$_2$Sb$_2$O band structure. (a) and (b), the photoemission intensity map at 150 K and 7 K respectively, The intensity was integrated over a window ($E_F$-10 meV, $E_F$+10 meV). (c), The symmetrized photoemission intensity along Γ-X direction at 150 K and 7 K, respectively. (d) and (e), Temperature dependence of the EDCs and symmetrized EDCs at the Fermi crossing of M, the EDCs are integrated over the red rectangle area in (c). (f), The temperature dependence of the CDW gap. The solid line is the fit to a mean field formula: $\Delta_0\sqrt{1-(T/T_{CDW})}$, where $\Delta_0$=65 meV, $T_{CDW}$=115 K. (g), Temperature dependence of the EDCs around the Fermi crossing near Γ, the EDCs are integrated over the yellow rectangle area in (c). Data were measured with 21.2 eV photons from a Helium discharge lamp.

The detailed temperature dependence of the low energy electronic structure of Na$_2$Ti$_2$Sb$_2$O is presented in Fig.4. The Fermi surface topologies of Na$_2$Ti$_2$Sb$_2$O at 150 K and 7 K are rather similar, but a dramatic spectra weight change can be observed around the X point. At 150 K, which is above the phase transition temperature at 115 K, the spectra weight around X is quite strong compared with those around the Γ point. At 7 K, which is well below the transition, the spectral weight near X is obviously suppressed, while it was slightly enhanced near Γ. Fig.4(c) shows the symmetrized spectrum along Γ-X. The band dispersion shows much alike at both temperatures, but an energy gap opens at X point when it comes into the CDW/SDW state at 7 K. We tracked the EDCs at the Fermi crossing of α band to reveal the CDW/SDW gap opening behavior more precisely. The density of states near $E_F$ is obviously suppressed with decreasing temperature; an energy gap opens at 113 K below the phase transition temperature of 115 K for Na$_2$Ti$_2$Sb$_2$O. The gap size increased with decreasing temperature, following the typical BCS formula[Fig.4(f)]. The gap size get saturated at low temperature and the largest gap size is about 65 meV at 7 K, which give a large ratio of 2Δ/k$_B$T$_s$~13. The optical study[24] revealed 2Δ/k$_B$T$_s$ ~14, in consistent with our findings. Such a large ratio indicates that this density wave system is in the strong electron–photon coupling regime[27].

Intriguingly, the photoemission spectrum of the electron band β around Γ shows a broad line shape without a sharp quasiparticle peak near $E_F$, and the spectral weight increases slightly with deceasing temperature [Fig.4(g)]. Furthermore, the peak position moves slightly upward to $E_F$ with deceasing temperature. Compared with the obvious gap opening behavior at X, it is safe to conclude that the gap does not open near Γ. In consideration of the theoretic prediction that X and M show multiple parallel sections, it is nature to deduce that Fermi surface nesting may happen between the parallels sections of X and M. Due to the Matrix element effects, the spectral weight near M is extremely weak for data taken with 21.2 eV photons, we thus cannot access the temperature dependence there. On the other hand, the large energy scale spectral weight transfer and broad lineshape, together with the large gap all suggest that the system is likely in the strong-coupling regime, therefore as previously observed in BaTi$_2$As$_2$O(Ref. 27), 2H-TaS$_2$(Ref. 26) and 2H-NbSe$_2$(Ref. 28), the low energy states over the entire Brillouin zone instead of only Fermi surface might be responsible for the density wave formation.

**Discussion**
It is crucial to understand the nature of the phase transition in the parent compounds of the newly discovered titanium-based oxypnictide superconductors, which is an essential step towards a thorough understanding of their superconducting mechanism. The SDW origin of the instability would favor an unconventional superconductivity with a possibly sign-changing *s*-wave pairing, while the CDW origin would suggest more conventional superconductivity with a simple *s*-wave pairing. Previous experimental and theoretical studies have evoked much controversy on the nature of the possible density wave transition. Our photoemission results are consistent with the density wave origin of the phase transition in $Na_2Ti_2Sb_2O$. Moreover, considering the qualitative agreement of the experimental results and the calculated electronic structure[23] in the nonmagnetic states, and it is reasonable to deduce that it is possibly a conventional CDW transition in $Na_2Ti_2Sb_2O$. Although further low temperature ARPES or STM experiment is certainly needed to reveal the exact nature of the superconducting samples, one can speculate that the superconductivity in $Na_xBa_{1-x}Ti_2Sb_2O$(Ref. 13) is likely due to electron phonon interactions, just like in $NbSe_2$(Ref. 28).

In summary, our experimental band structure agrees qualitatively well with the calculation[23] in the nonmagnetic state, excluding the existence of possible magnetic order in $Na_2Ti_2Sb_2O$. $Na_2Ti_2Sb_2O$ shows obvious multi-band and multi orbital nature, which resemble the iron-based superconductors. The electron band at M and the hole band at X show weak $k_z$ dispersion, consistent with its layered crystal structure. We observe a large density wave gap of 65 meV which forms near the X point at 7 K, indicating that $Na_2Ti_2Sb_2O$ is likely a CDW material. The weak renormalization of the overall band structure indicates weak electron-electron correlation, while the broad lineshape and large energy gap and spectral weight transfer suggest the system is likely in the strong electron-phonon interaction regime.

## Methods

**Sample synthesis.** Single crystals of $Na_2Ti_2Sb_2O$ were synthesized by the self-flux method. A mixture of Na, Sb, Ti and $Ti_2O_3$ with molar ratio of 18:18:1:4 is prepared and put into an aluminum oxide crucible sealed inside a Ta tube. The mixture is gradually heated to 800 °C and quenched to room temperature. Afterwards the mixture is heated at 1100 °C for 2 hours and cooled to 500 °C at 5 °C/hour before quenched to room temperature.

**ARPES measurement**. The polarization and photon energy dependent ARPES data were taken at the surface and interface spectroscopy beamline of the Swiss Light Source (SLS), The temperature dependent ARPES data were taken with an in-house setup at Fudan University. All data were collected with Scienta R4000 electron analyzers. The overall energy resolution was 15 meV or better, and the typical angular resolution was 0.3°. The samples were cleaved *in-situ* and measured under ultrahigh vacuum better than $3\times10^{-11}$ mbar.

## Acknowledgements
We thank Dr. M. Shi for the experimental support at Swiss Light Source(SLS). We gratefully acknowledge helpful discussions with Prof. N. L. Wang. This work is supported in part by the National Science Foundation of China and the National Basic Research Program of China (973 Program) under Grants No. 2012CB921400, No. 2011CB921802, No.2011CBA00112, No. 2011CB309703, and No. 91026016. The single crystal growth work at the University of Tennessee was supported by the U.S. DOE, BES, through Contract No. DE-FG02-05ER46202.

## Author contributions
S. Y. Tan, J. Jiang, Z. R. Ye, X. H. Niu and B. P. Xie performed the ARPES measurements. Y. Song, C. L. Zhang and P. C. Dai provided the single crystal samples. S. Y. Tan and D. L. Feng analyzed the ARPES data, S. Y. Tan and D. L. Feng wrote the paper. D. L. Feng and X. C. Lai are responsible for the infrastrure, projection and planning.

## Additional information
Competing financial interests: The authors declare no competing financial interests. Correspondence and requests for materials should be addressed to D.L.Feng (dlfeng@fudan.edu.cn)